\documentclass{aa}

\usepackage{graphicx}
\usepackage{xcolor}
\usepackage{txfonts}
\usepackage{gensymb}
\usepackage{textgreek}
\usepackage{color}
\usepackage[colorlinks=true]{hyperref}
\hypersetup{linkcolor=blue,citecolor=blue,filecolor=blue,urlcolor=blue}

\begin{document} 

   \title{Polarisation angle variability in tidal disruption events}

\author{A. Floris \inst{1,2,3}
  \and I. Liodakis \inst{1}
  \and K. I. I. Koljonen \inst{4}
  \and E. Lindfors \inst{5,6}
  \and B. Ag\'is-Gonz\'alez\inst{1}
  \and A. Paggi \inst{1,2}
  \and D. Blinov \inst{1,2}
  \and K. Nilsson \inst{6}
  \and I. Agudo \inst{7}
  \and P. Charalampopoulos \inst{6}
  \and J. Escudero Pedrosa \inst{7,8}
  \and V. Piirola \inst{5}
}

    \institute{Institute of Astrophysics, FORTH, N.Plastira 100, Vassilika Vouton, 70013 Heraklion, Greece\\  \email{afloris@ia.forth.gr}
    \and Department of Physics University of Crete, Voutes University Campus, 70013 Heraklion, Greece
    \and National Institute for Astrophysics (INAF), Astronomical Observatory of Padova, IT-35122 Padova, Italy
    \and Institutt for Fysikk, Norwegian University of Science and Technology, Høgskloreringen 5, Trondheim, 7491, Norway
    \and Department of Physics and Astronomy, 20014 University of Turku, Finland
    \and Finnish Centre for Astronomy with ESO (FINCA), Quantum, Vesilinnantie 5, 20014 University of Turku, Finland
    \and Instituto de Astrofísica de Andalucía, IAA-CSIC, Glorieta de la Astronomía s/n, 18008 Granada, Spain
    \and Center for Astrophysics \textbar ~Harvard \& Smithsonian, 60 Garden Street, Cambridge, MA 02138 USA
    }
   \date{}

  \abstract
   {Tidal disruption events (TDEs) occur when a star is disrupted by the tidal forces of a supermassive black hole, and these events produce bright multi-wavelength flares. Polarimetric measurements of TDEs allow us to disentangle the geometry and the mechanisms characterising the accretion process.}
   {We carried out the first systematic study of the time evolution of the optical polarisation angle ($\Theta$) in a sample of classified TDEs, combining our own data with all available measurements from the literature, with the goal of testing the currently available models that describe TDE emission.}
   {We assembled data from all available observing epochs with significant linear polarisation detections (\(\Pi-3\sigma_\Pi>0\%\)) for sources with at least two such epochs, and we determined the overall variability trends across the sample in various time frames, such as days from peak time and the fallback time ($t_0$) derived from the different models.}
   {Our final sample comprises 12 transients, including three Bowen fluorescence flares (BFFs). The majority of the sources show significant $\Theta$ variability. The distribution of $|\mathrm{d}\Theta/\mathrm{d}t|$ peaks near \(\sim2^{\circ}\,\mathrm{d}^{-1}\). BFFs tend to display sustained late-time $\Theta$ evolution, likely due in part to their slower fading. No universal trend emerges when time is normalised by $t_0$.}
   {Short-timescale $\Theta$ variability is common in TDEs and is difficult to reconcile with simple axisymmetric reprocessing models that predict a constant polarisation angle. The observed phenomenology favours scenarios with evolving, non-axisymmetric geometries and/or shocks, possibly coupled with changes in optical depth. Denser polarimetric monitoring, contemporaneous spectroscopy, and X-ray/UV coverage are required to break the remaining degeneracies.}


   \authorrunning{A. Floris et al.}
   \titlerunning{Polarisation angle variability in TDEs}
   \maketitle

\section{Introduction}
\label{intro}

Tidal disruption events (TDEs) are luminous transients that occur when a star approaches a supermassive black hole (SMBH) close enough for the tidal field of the SMBH to overwhelm the star’s self-gravity, shredding the star and liberating a large fraction of its gravitational binding energy \citep{hills1975,rees1988,komossa2015}. The ensuing flare radiates across the X-ray, ultraviolet (UV), optical, and infrared (IR) bands, yet the mechanism that powers this emission remains unclear. Delayed radio flares have also been observed in several sources \citep{alexander2025}

Two broad families of models have been proposed. In the reprocessing scenario, a compact accretion disc quickly forms and emits soft X-rays and extreme-UV photons that are thermalised in an optically thick, quasi-spherical envelope at radii far larger than the tidal radius, producing the near UV–optical continuum \citep{metzger2016,dai2018}. In contrast, shock-powered models attribute the luminosity to kinetic energy released when the highly eccentric debris streams self-intersect at large apocentric distances, converting orbital energy into heat and radiation \citep{piran2015,shiokawa2015}. There is observational evidence that supports both pictures: narrow line emission triggered by resonant He\textsc{ii} and Lyman~$\alpha$ pumping of O\textsc{iii]} and N\textsc{iii} (Bowen fluorescence) and forbidden transitions from highly ionised species that require a hard ionising continuum (coronal lines, e.g.\ [Fe\textsc{x}] and [Fe \textsc{xiv}]; \citealt{leloudas2019,blagorodnova2019,trakhtenbrot2019,koljonen2024}) point to a buried X-ray source, whereas the rapidly variable polarisation reported in several TDEs \citep{koljonen2024,koljonen2025,floris2025}, as well as high-degree optical polarisation (e.g.\ $\Pi\sim25\%$ in \object{AT~2020mot}; \citealt{liodakis2023}) and X-ray brightening events after the optical-UV peak \citep{kajava2020}, point to an aspherical, evolving scattering- or shock-dominated geometry, and possibly ordered magnetic fields. However, late X-ray brightenings have also been interpreted as a reprocessing layer becoming optically thin \citep[e.g.][]{guolo2024,reynolds2025,earl2025}.

Polarimetry is uniquely sensitive to the geometry and magnetic-field structure and can potentially solve this ambiguity. In the rare class of jetted TDEs, the synchrotron emission from the jet should be intrinsically polarised. The first jetted-TDE observations found a $\Pi$ of $\sim 7-8\%$ near the peak that decreased as the events faded \citep[the first one being a $2\sigma$ detection;][]{wiersema2012,wiersema2020}. However, the most recent jetted TDE, AT2022cmc, exhibited no significant polarisation in the the rest-frame UV when observed approximately 10 days after peak brightness \citep{cikota2023}. In optical TDEs with no evidence of jet emission, the linear polarisation is expected to arise predominantly from electron scattering. The polarisation degree ($\Pi$) measures the departure from spherical symmetry, while the polarisation angle ($\Theta$) traces the projected orientation of the scattering geometry---or, when an ordered magnetic field is present, the field's direction itself. Systematic changes in $\Theta$ have long been recognised as key diagnostics in other compact-object systems, for example: 

\begin{itemize}
    \item Blazars: Smooth, large-amplitude $\Theta$ of $>90^{\circ}$ typically occur on timescales of days to weeks and are frequently contemporaneous with gamma-ray flares \citep[e.g.][]{blinov2015,kiehlmann2016,agudo2018}.

    \item Gamma-ray bursts (GRBs): Afterglow monitoring of GRB~091018 and similar events has revealed $\sim90^{\circ}$ swings in $\Theta$ near the jet-break time, consistent with geometric models of off-axis expanding jets \citep[e.g.][]{wiersema2012b}.

    \item X-ray binaries (XRBs): In the XRB source 4U~0115+63, $\Theta$ evolved by $\sim20^{\circ}$ over a single outburst, tracing the precession of a warped accretion disc \citep[e.g.][]{reig2018}. Rapid, lower-amplitude changes---and occasional polarised flares---were also reported in the black-hole transient V404Cyg during its 2015 outburst \citep[e.g.][]{shahbaz2016}.

    \item Magnetars: Observations of 4U~0142+61 with the Imaging X-ray Polarimetry Explorer (IXPE) uncovered an energy-dependent $90^{\circ}$ rotation of the X-ray polarisation angle at 4-5 keV, potentially due to quantum-electrodynamic mode conversion in the magnetised atmosphere \citep{taverna2022}.
\end{itemize}

Analogous behaviour in TDEs could offer a direct probe of disc formation, stream dynamics, and magnetic-field topology. Indeed, coordinated variability in $\Pi$ and $\Theta$ has been reported for AT2020mot and AT2023clx \citep{koljonen2025} when time is expressed in units of fallback time ($t_0$), pointing to the possibility of a common evolution scenario.

Additionally, recent spectropolarimetry of the nearby TDE AT2023clx supports the hypothesis that the observed polarisation arises from scattering in a dusty torus around the SMBH \citep{uno2025}, although alternative interpretations involving shocks have also been proposed \citep{koljonen2025}. The latter scenario would most naturally apply to events occurring within a pre-existing active galactic nucleus. Seyfert galaxies, for example, exhibit electron-scattering-dominated polarisation and a bimodal distribution of $\Pi$, with Seyfert~1 galaxies showing systematically lower polarisation than Seyfert~2 galaxies \citep{hutsemekers2017}. The polarisation angle, however, is generally found to remain approximately constant in the majority of studied sources, even when $\Pi$ varies over time, in both optical and X-ray polarimetry \citep{antonucci1982,gaskell2012,marin2017,marin2024}.

During our dedicated time-domain polarimetric campaign targeting TDEs, the Black hOle Optical polarization TimE-domain Survey (BOOTES), we found that most events are either consistent with 0\% polarisation within the errors or show $\Pi\lesssim6\%$ at single epochs, and are thus still compatible with both reprocessing and shock-powered scenarios \citep{floris2025}. However, the discovery of $\Theta$ variability and $\Pi$ excursions well above the $\lesssim6\%$ limit \citep{liodakis2023,koljonen2024,koljonen2025,floris2025} expected for pure electron-scattering envelopes \citep{leloudas2022} suggests that, at least in some TDEs, the scattering region is highly aspherical, dynamically evolving, and possibly threaded by ordered magnetic fields.

Motivated by these findings, we present the first systematic study of the temporal evolution of $\Theta$ in a sample of optically selected TDEs observed as part of BOOTES and supplemented with all publicly available data from the literature.
In Sect. \ref{sample} we present the source sample used in this work, together with new polarimetric observations. The results are presented in Sect. \ref{res}, and the implications derived from them are discussed in Sect. \ref{discussions}. We provide a summary of our conclusions in Sect. \ref{conclusions}. 
Throughout this paper we adopt a flat $\Lambda$ cold dark matter cosmology with $H_0 = 67.4\ \mathrm{km\ s^{-1}\ Mpc^{-1}}$, $\Omega_\mathrm{M}=0.315$, and $\Omega_\Lambda=0.685$  \citep{planck2020}.

\section{Sample}
\label{sample}

The sample used in this work consists of the observations presented in \cite{floris2025} and taken as part of the BOOTES observation campaign, complemented with new observations of AT2024pvu and by all literature measurements with at least two significant detections of $\Pi$ (i.e. $\Pi-3\sigma_\Pi>0\%$), ensuring reliable $\Theta$ estimates and temporal tracking. We included all literature data satisfying this criterion in our analysis; these measurements were drawn from \cite{leloudas2022}, \cite{koljonen2024,koljonen2025}, \cite{jordanamitjans2025}, and \cite{floris2025}. We re-analysed Nordic Optical Telescope (NOT) data for AT2019dsg from \cite{lee2020} and obtained lower polarisation values, which we adopted in this work. The data were analysed using the semi-automatic pipeline for NOT polarimetric data \citep{hovatta2016}, which has been repeatedly tested on faint targets, including TDEs \citep{koljonen2024,koljonen2025,floris2025}. We also included imaging polarimetry data from Very Large Telescope (VLT) archival observations of TDEs AT2020zso and AT2021blz. The final sample comprises 12 sources, three of which (AT2019aalc, AT2020afhd, and AT2022fpx) have also been suggested to be Bowen fluorescence flares (BFFs) due to prominent Bowen lines.

To maximise temporal coverage, we combined measurements across filters after verifying the $\Theta$  values are consistent between bands; for electron-scattering, $\Theta$ is expected to be approximately achromatic at a fixed geometry. To reduce noise and redundancy, we adopted a minimum cadence of one day and retained the most precise measurement per epoch (typically $R$ band for NOT, $V$ band for VLT, and $L$ band for the Liverpool Telescope (LT) from \citealt{jordanamitjans2025}).

We computed $|\Delta\Theta/\Delta t|$ between adjacent detections and constructed per-source histograms. Where needed, we unwrapped the $180^\circ$ ambiguity by requiring continuity in $(q,u)$ space. All reported uncertainties are propagated from the Stokes parameter errors. We report additional observations from the BOOTES campaign, VLT, and re-analysed NOT data from \cite{lee2020} in Table \ref{tab:newpol}. The measurements have been processed similarly to  \cite{floris2025} but were not host-corrected. As a matter of fact, in the assumption of an unpolarised host galaxy, $\Theta$ is not expected to change during host correction \citep{floris2025}. While this work was under review, additional TDE polarimetric observations were published in \cite{wichern2026}, including of AT2020zso and AT2021blz, which we also present in Table \ref{tab:newpol}. These data were not utilised in our analysis; however, for these observations we verified that our measurements and those reported in \cite{wichern2026} are consistent within \(3\sigma\).

\begin{table}[h!]
\renewcommand{\arraystretch}{1.25}
\caption{TDE peak times.}
\label{tab:tpeak}
\centering
\begin{tabular}{c c | c c}
\hline\hline
Name & $t_{\rm peak}$ & Name & $t_{\rm peak}$\\
 & [d] & & [d]\\
 (1) & (2) & (3) & (4) \\
\hline
AT2018dyb & 58340.18 & AT2021blz & 59259.23 \\
AT2019aalc & 60139.60 & AT2022fpx & 59785.53\\
AT2019dsg & 58607.47 & AT2023clx & 59987.50\\
AT2020mot & 59057.87 & AT2023lli & 60169.69\\
AT2020zso & 59191.85 & AT2024gre & 60441.22\\
AT2020afhd & 60351.53 & AT2024pvu & 60544.15\\
\hline
\end{tabular}
\tablefoot{(1) and (3): TDE name. (2) and (4): Peak time (MJD).}
\end{table}

\section{Results}
\label{res}

\begin{figure}
\centering
\includegraphics[width=\hsize]{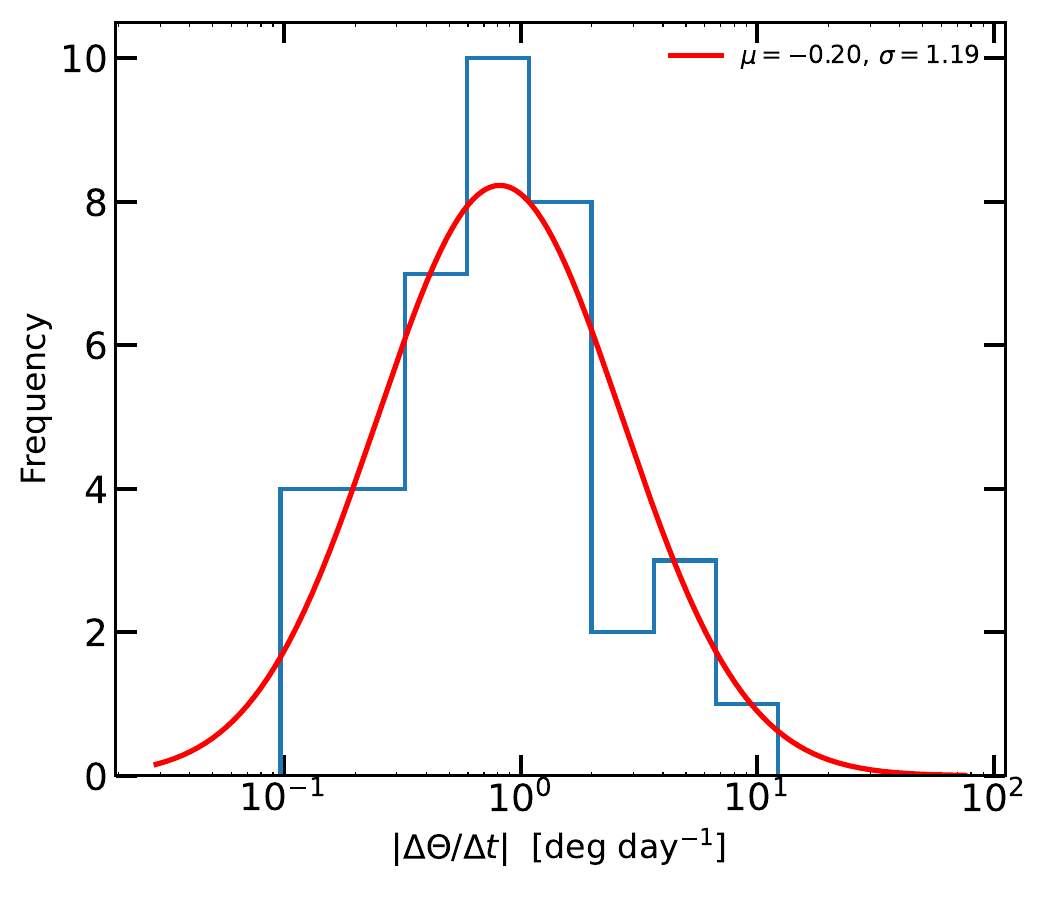}
\caption{Histogram of the rate of change of the polarisation angle per day for the sources in the sample. Non-detections are not considered in this calculation, as $\Theta$ is undefined in such cases. In red, a log-normal distribution defined in base $e$ with the same central value ($\mu$) and dispersion ($\sigma$) of the data is overplotted.}
\label{fig:histrate}
\end{figure}

\begin{figure}
\centering
\includegraphics[width=\hsize]{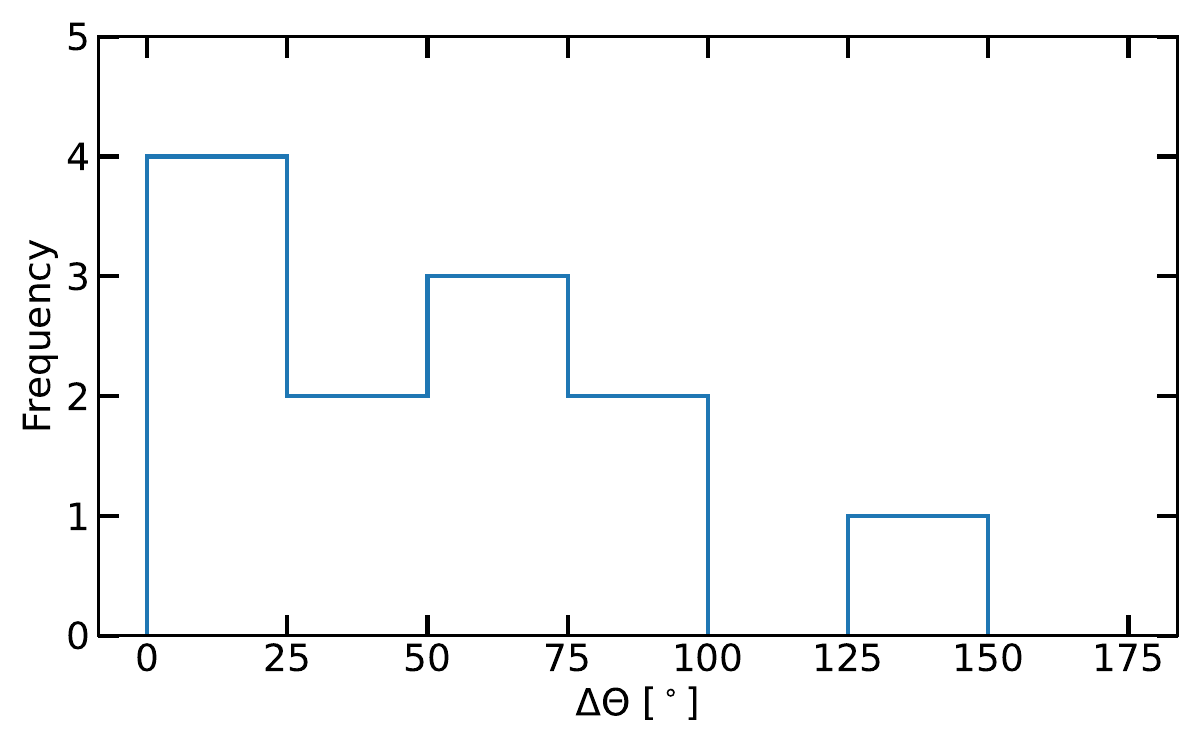}
\caption{Maximum $\Theta$ excursion per source in $25^\circ$ bins. Non-detections are omitted, as $\Theta$ is undefined in such cases.}
\label{fig:amplitudechange}
\end{figure}

In this work we quantified $\Theta$ variability via adjacent-epoch rates and per-object amplitudes, selecting the measurements by adopting the methodology described in Sect. \ref{sample}.
We estimated the rate of change of the polarisation angle ($|\Delta\Theta/\Delta t|$) in a manner analogous to studies of blazars \citep{blinov2016}, providing a benchmark against which future models can be tested. At present, no model predicts a specific $|\Delta\Theta/\Delta t|$, with the exception of an axisymmetric reprocessing scenario that implies $|\Delta\Theta/\Delta t|\sim 0 ^\circ {\rm d}^{-1}$ on short timescales.
Figure~\ref{fig:histrate} shows that the distribution of $|\Delta\Theta/\Delta t|$ follows a log-normal distribution that peaks  at $1.90^\circ$ d$^{-1}$, with a standard deviation of $\sigma =3.78^\circ$ d$^{-1}$ (shown as the red curve overplotted on the histogram). We verified this assumption with the Kolmogorov-Smirnov and Anderson-Darling tests, obtaining a p-value of 0.95 (i.e. both tests show that we cannot reject the null hypothesis that the distributions are drawn from a log-normal distribution). We have overplotted a log-normal distribution with the same mean and standard deviation to guide the eye. Additionally, for each pair of consecutive $\Delta\Theta$ measurements, we performed a Monte Carlo resampling within the corresponding uncertainties (10{,}000 realisations) to assess whether the uncertainties are underestimated, which would affect our variability estimates. This procedure yielded a log-normal distribution centred at $|\Delta\Theta/\Delta t|=2.09^\circ$ d$^{-1}$ with a standard deviation $\sigma=4.28^\circ$ d$^{-1}$, consistent with the observed distribution. To show the rate of occurrence and the magnitude of $\Theta$ variability in our sample, we also display the maximum absolute amplitude of the $\Theta$ variation for each source in Fig. \ref{fig:amplitudechange}.

\begin{figure}
\centering
\includegraphics[width=\hsize]{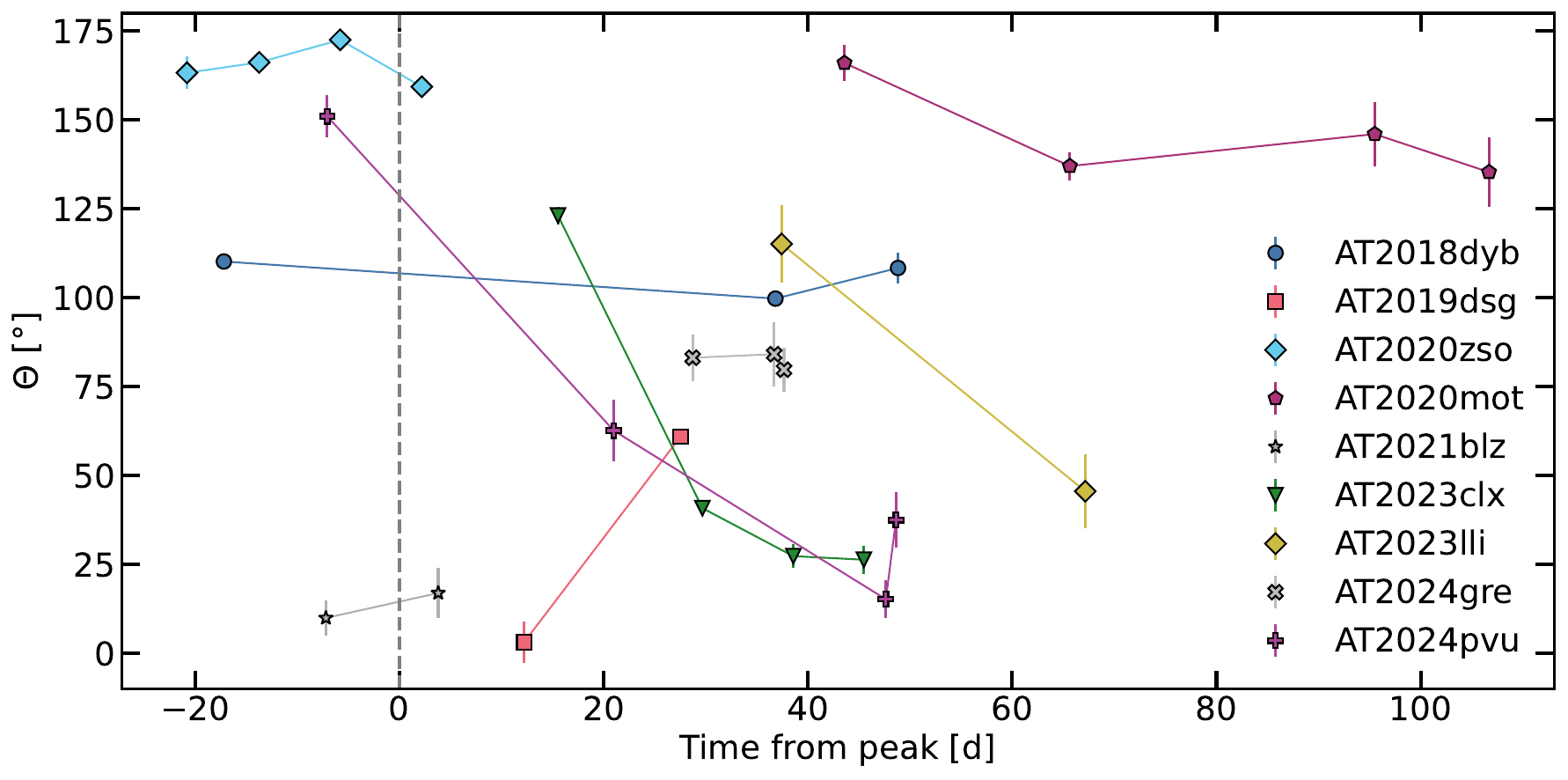}
\caption{$\Theta$ as a function of days since optical peak for the TDEs in the sample. Non-detections are omitted, as $\Theta$ is undefined in such cases.}
\label{fig:peaktimetde}
\end{figure}

Most objects are variable, some with small-amplitude variabilities ($\Delta\Theta \lesssim 25^\circ$) and some with large-amplitude variations ($\Delta\Theta \gtrsim 90^\circ$).
We present plots of $\Theta$ as a function of time from the peak (reported in Table \ref{tab:tpeak} for each source) for TDEs and BFFs in Figs.~\ref{fig:peaktimetde} and \ref{fig:peaktimebff}, respectively. Alternatively, we consider the variability of $\Theta$ as a function of fallback time ($t_0$) as in \cite{koljonen2025} in Appendix \ref{polanglefig}. However, expressing time in units of $t_0$ does not reveal a common trend.

Additionally, similarly to \cite{floris2025}, we studied the rate occurrence of variability in our sample. Of the eight sources for which we detect significant variability between two consecutive observations, three of them (AT2019dsg, AT2020zso, and AT2023lli) show minimum variability timescales comparable to the minimum observational gap, indicating that denser sampling could reveal faster changes in some TDEs. High-S/N observations (e.g. with VLT for AT2020zso) enable the detection of significant but relatively small-amplitude variations on short (approximately weekly) timescales. AT2020zso falls in the $0$-$25^{\circ}$ bin of Fig.~\ref{fig:amplitudechange}.

\subsection{TDE behaviour}

From our analysis, we see that, even in the case of limited observations, there is significant variation in $\Theta$ in the majority of the TDEs in our sample. The main exceptions are  AT2021blz and AT2024gre: the former likely because only two observations spanning $\sim11$ days were available, the latter because the observations span only $\sim 9$ days.  Nevertheless, both cases are broadly consistent with both reprocessing models (which predict no changes in $\Theta$ over short timescales; \citealt{metzger2016,leloudas2022}) and with the shock model that predicts changes in $\Theta$ after $t>t_0$ \citep{piran2015,liodakis2023}. By contrast, AT2024pvu exhibits an overall amplitude of $\sim 135^\circ$ from the first observation (obtained a few days before the peak) and the last two observations ( $\sim45$d later). Such large continuous smooth rotations of the polarisation angle have so far only been observed in blazar jets \citep[e.g.][]{blinov2018,liodakis2020}. The source displays a steady rotation at a rate of $\sim3^\circ $d$^{-1}$. Several other TDEs show qualitatively similar, though smaller, variations. Overall, we find that most TDEs exhibit variability in $\Theta$, which in several cases diminishes with time as $\Theta$ approaches an approximately constant value (e.g. AT2020mot and AT2023clx). This inference is necessarily limited by the cadence and duration of the available monitoring: for sources such as AT2020zso or AT2024pvu, additional late-time observations could have revealed whether a similar settling behaviour occurs after the final epochs in our dataset.

\subsection{Bowen fluorescence flare behaviour}

\begin{figure}
\centering
\includegraphics[width=\hsize]{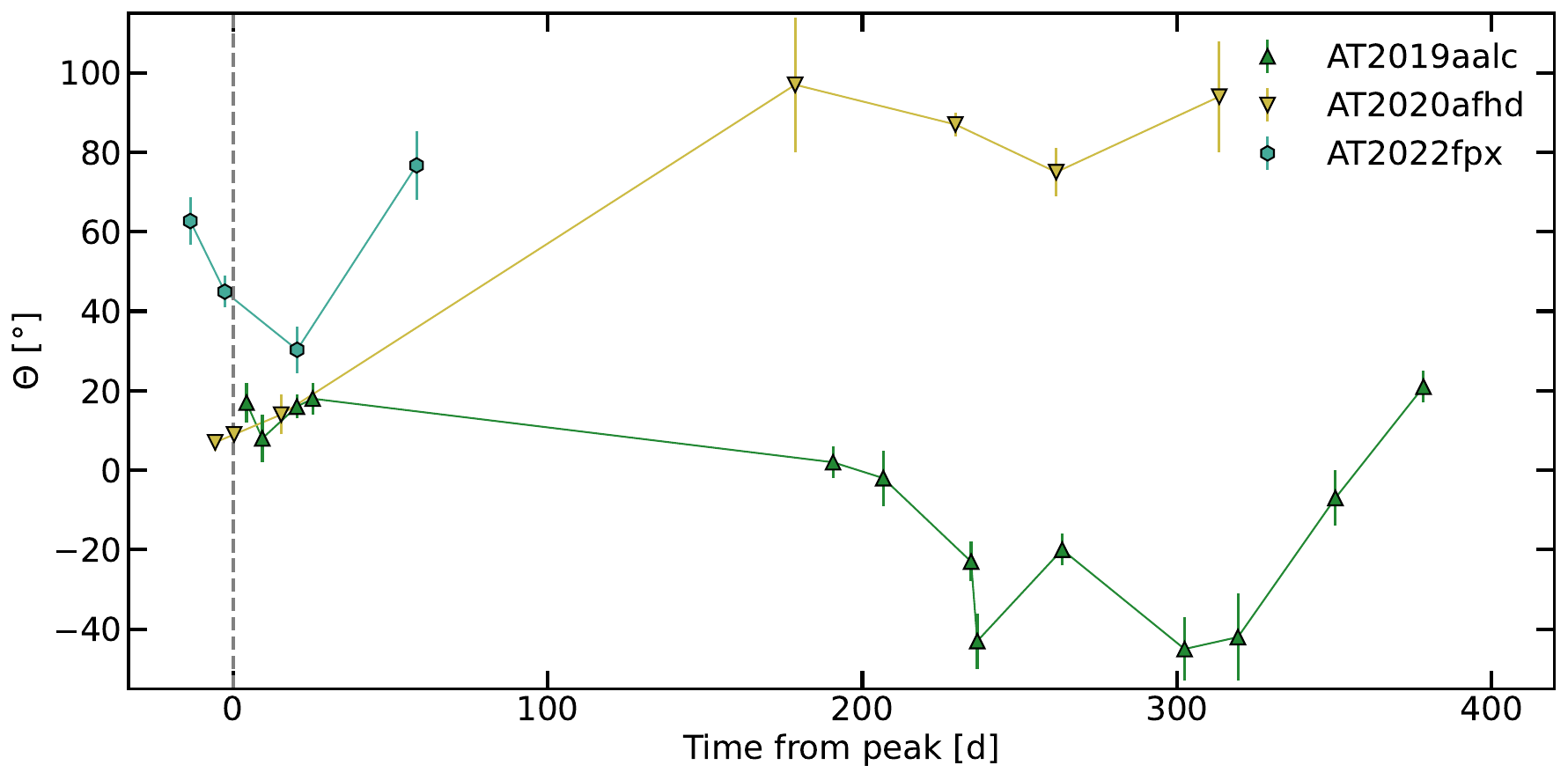}
\caption{$\Theta$ as a function of days since optical peak for the BFF subsample. Non-detections are omitted, as $\Theta$ is undefined in such cases.}
\label{fig:peaktimebff}
\end{figure}

As mentioned in Sect. \ref{sample}, in this work we studied three BFFs. AT2022fpx and AT2020afhd are also considered TDEs \citep{perez2022,koljonen2024,hammerstein2024}, while AT2019aalc is considered to be associated with a nuclear transient and possibly coupled with a pre-existing active galactic nucleus disc \citep{sniegowska2025, jordanamitjans2025}; this likely gives rise to at least part of the observed variability. We show in Fig. \ref{fig:peaktimebff} that, similarly to the TDEs in our sample, the BFFs exhibit variability in $\Theta$. BFFs, which typically change more slowly than TDEs, tend to have later-time detections, which enables a better sampling of their $\Theta$ evolution. AT2022fpx displays an initial gradual decrease followed by a rapid increase, after which the polarisation degree becomes consistent with 0\% within the errors \citep{koljonen2025}. AT2019aalc and AT2020afhd exhibit sharp changes on approximately weekly timescales at late times; AT2020afhd maintains a quasi-constant $\Theta$ of $\sim15^{\circ}$ early on and transitions to a distinct value ($\sim100^\circ$) after a \(\sim6\)-month gap. Figure \ref{fig:peaktimebff} suggests that the BFFs AT2019aalc and AT2022fpx undergo changes in the direction of $\Theta$ rotation over the course of the flare, whereas none of the classical TDEs in our sample shows comparable behaviour. However, given the small number of objects in the TDE and BFF subsamples, we cannot draw robust conclusions about the physical origin of these changes.

\section{Discussion}
\label{discussions}

Figure \ref{fig:amplitudechange} demonstrates that $\Theta$ variability is common in our sample, with some TDEs showing small (\(<25^{\circ}\)) and others very large (\(\gtrsim90^{\circ}\)) excursions over the observed duration. We identified several phenomenological patterns and predict how to identify them in the future:

\begin{itemize}
\item Stochastic or chaotic changes in $\Theta$: The shock-powered framework \citep{piran2015, shiokawa2015} naturally accommodates irregular $\Pi$ and $\Theta$ changes as the location, orientation, and strength of the tidal shocks evolve. This behaviour is consistent with the late-time variability of AT2019aalc and with the joint $\Pi$-$\Theta$ evolution observed in AT2020mot, AT2022fpx, and AT2023clx \citep{liodakis2023, koljonen2024, koljonen2025}. The $\Theta$ changes seen in AT2019dsg, AT2023lli, and AT2024pvu can likewise be attributed to this scenario.
\item Near-constant $\Theta$: In cases with little evolution (e.g. AT2024gre and AT2021blz, even though these results are biased because of the very short time span of the observations), simple axisymmetric reprocessing remains viable; modest deviations could reflect second-order asymmetries \citep{leloudas2022}.
\item Opacity and obscuration changes: AT2020afhd appears to transition between distinct $\Theta$ states after \(\sim6\)~months. This may indicate a change from an optically thick to an optically thin reprocessing layer or a modification of the shock geometry as the debris circularises, potentially accompanied by the emergence of accretion-disc emission. A hybrid scenario with both shocks and reprocessing layers \citep[e.g.][]{charalampopoulos2023} could also reproduce such behaviour. Concomitant spectral changes (e.g. the emergence of He\textsc{ii}$\lambda4686$ once the gas becomes optically thin) and/or the appearance or strengthening of X-rays would be expected; in AT2020afhd, stronger late-time X-rays are indeed present \citep{floris2025}. A related possibility is suggested by the X-ray behaviour of AT2022fpx: during the X-ray flare, the optical polarisation was $<0.5$\% \citep{koljonen2024}, likely due to a combination of lower luminosity and the depolarised nature of accretion disc emission. Had the S/N been higher, contemporaneous $\Theta$ variations analogous to those observed in AT2020afhd might have been measurable.
\item Polar-equatorial emission switch: The $\sim90^\circ$ change in $\Theta$ during the evolution of AT2020afhd may indicate a transition between polar and equatorial scattering regions, as predicted for active galactic nuclei \citep{marin2017,marin2020}.
\item Periodicity: Within our sample, $\Theta$ does not show signs of periodicity. Nevertheless, two main instances predict periodic changes in $\Pi$ and $\Theta$. Accretion disc precession has been invoked to explain the observed X-ray behaviour of TDE AT2020ocn \citep{pasham2024}, and it is expected to also manifest in periodic changes in polarisation when the orbital plane of the disrupted star and the spin axis of the black hole are misaligned \citep{teboul2023,koljonen2024}. Another interesting possibility is a TDE occurring in a SMBH binary system \citep{lopez2019}, which can likewise produce periodic variations in both $\Pi$ and $\Theta$ \citep{dotti2022,marin2023}.
\end{itemize}

Overall, the observations are difficult to reconcile with purely axisymmetric, static reprocessing. Non-axisymmetric and time-dependent geometries (shock- or wind-driven), together with evolving optical depth, provide a more natural explanation and offer testable predictions.

\section{Conclusions}
\label{conclusions}

In this work we analysed all currently available optical polarimetric datasets with at least two significant polarisation angle measurements from the literature and from the BOOTES campaign \citep{floris2025}. We draw several key conclusions:

\begin{itemize}
    \item Overall $\Theta$ variability ($\Delta\Theta>25^\circ$) is observed in 8 out of 12 sources. The 4 sources with a constant $\Theta$ have been observed only over a short time span (e.g. AT2024gre, 9 days).

    \item The distribution of $|\Delta\Theta/\Delta t|$ peaks near $\sim2^{\circ} \mathrm{d}^{-1}$.

    \item Based on an exclusion criterion, we find that simple axisymmetric reprocessing cannot explain the ensemble. Models invoking shocks, evolving optical depth or geometry, and/or multiple polarising components better reproduce the observed phenomenology.

    \item BFFs tend to exhibit sustained late-time $\Theta$ evolution. This is likely due in part to their slower fading, which in turn is due to their systematically longer decay times allowing for higher-S/N measurements to be taken over a longer period of time.

    \item Denser polarimetric monitoring, contemporaneous spectroscopy, and X-ray/UV coverage are needed to discriminate among models.

    \item These results demonstrate the need for predictive, time-dependent polarimetric modelling (including radiative-transfer and polarisation simulations) that can reproduce $\Theta(t)$ and be directly tested against current and future observations.
\end{itemize}

\begin{acknowledgements}

AF, IL, AP and BAG were funded by the European Union ERC-2022-STG - BOOTES - 101076343. Views and opinions expressed are however those of the author(s) only and do not necessarily reflect those of the European Union or the European Research Council Executive Agency. DB acknowledges support from the European Research Council (ERC) under the Horizon ERC Grants 2021 programme under grant agreement No. 101040021. KIIK has received funding from the European Research Council (ERC) under the European Union’s Horizon 2020 research and innovation programme (grant agreement No. 101002352, PI: M. Linares). Neither the European Union nor the granting authority can be held responsible for them. The IAA-CSIC co-authors acknowledge financial support from the Spanish "Ministerio de Ciencia e Innovaci\'{o}n" (MCIN/AEI/ 10.13039/501100011033) through the Center of Excellence Severo Ochoa award for the Instituto de Astrof\'{i}sica de Andaluc\'{i}a-CSIC (CEX2021-001131-S), and through grants PID2019-107847RB-C44 and PID2022-139117NB-C44. The data in this study include observations made with the Nordic Optical Telescope, owned in collaboration by the University of Turku and Aarhus University, and operated jointly by Aarhus University, the University of Turku and the University of Oslo, representing Denmark, Finland and Norway, the University of Iceland and Stockholm University at the Observatorio del Roque de los Muchachos, La Palma, Spain, of the Instituto de Astrofisica de Canarias. The data presented here were obtained in part with ALFOSC, which is provided by the Instituto de Astrof\'{\i}sica de Andaluc\'{\i}a (IAA) under a joint agreement with the University of Copenhagen and NOT. Some of the data are based on observations collected at the Observatorio de Sierra Nevada; which is owned and operated by the Instituto de Astrof\'isica de Andaluc\'ia (IAA-CSIC). 
Based in part on observations collected at the Centro Astronómico Hispano en Andalucía (CAHA) at Calar Alto, operated jointly by Junta de Andalucía and Consejo Superior de Investigaciones Científicas (IAA-CSIC). This research is based in part on observations collected at the European Southern Observatory under ESO programme 106.214S.001. We acknowledge funding to support our NOT observations from the Finnish Centre for Astronomy with ESO (FINCA), University of Turku, Finland (Academy of Finland grant nr 306531). E.L. was supported by Academy of Finland projects 317636 and 320045. This research has made use of data from the RoboPol programme, a collaboration between Caltech, the University of Crete, IA-FORTH, IUCAA, the MPIfR, and the Nicolaus Copernicus University, which was conducted at Skinakas Observatory in Crete, Greece.

\end{acknowledgements}

\bibliographystyle{aa}
\bibliography{biblio.bib}

\begin{appendix}

\onecolumn
\section{New observations}
\label{newobs}

In this section we present the results from the polarimetric observations of the TDEs employed in this work in Table \ref{tab:newpol}, as discussed in Sect. \ref{sample}.

\begin{table*}[h!]
\renewcommand{\arraystretch}{1.10}
\caption{New polarisation observations and re-analysed archival data.}
\label{tab:newpol}
\centering
\begin{tabular}{c c c c c c c c c}
\hline\hline
Name & MJD & $\Delta t_{\rm peak}$ & $\Pi$ & $\Theta$ & q & u & Telescope & Filter  \\
 & [d] & [d] & [\%] & [$^\circ$] & [\%] & [\%] & & \\
(1) & (2) & (3) & (4) & (5) & (6) & (7) & (8) & (9)\\
\hline
AT2019dsg & 58620.188 & +16.29 & $<1.31$ & ... & $0.68\pm0.44$ & $0.58\pm0.43$ & NOT & V \\
 & 58654.150 & +50.25 & $<0.96$ & ... & $-0.50\pm0.32$ & $0.18\pm0.32$ & NOT & V \\
 & 58680.054 & +76.15 & $<2.19$ & ... & $-0.97\pm0.69$ & $-1.09\pm0.77$ & NOT & V \\
 & 58681.165 & +77.27 & $<2.14$ & ... & $-1.59\pm0.72$ & $-0.37\pm0.70$ & NOT & V \\
AT2020zso & 59171.041 & $-20.77$ & $1.30\pm0.18$ & $170.9\pm4.0$ & $1.24\pm0.18$ & $-0.41\pm0.18$ & VLT & V \\
& 59171.048 & $-20.76$ & $1.61\pm0.19$ & $179.6\pm3.4$ & $1.61\pm0.19$ & $-0.02\pm0.19$ & VLT & B \\
& 59171.057 & $-20.75$ & $1.02\pm0.16$ & $163.2\pm4.53$ & $0.85\pm0.16$ & $-0.56\pm0.16$ & VLT & R \\
& 59178.071 & $-13.72$ & $0.81\pm0.07$ & $169.3\pm2.6$ & $0.76\pm0.08$ & $-0.30\pm0.07$ & VLT & V \\
& 59178.089 & $-13.70$ & $1.25\pm0.06$ & $8.6\pm1.4$ & $1.19\pm0.06$ & $0.37\pm0.06$ & VLT & V \\
& 59178.106 & $-13.69$ & $1.69\pm0.10$ & $4.9\pm1.7$ & $1.67\pm0.10$ & $0.29\pm0.10$ & VLT & B \\
& 59178.127 & $-13.68$ & $1.10\pm0.07$ & $166.2\pm1.8$ & $0.97\pm0.07$ & $-0.51\pm0.07$ & VLT & R \\
& 59186.054 & $-5.76$ & $1.28\pm0.08$ & $172.5\pm1.7$ & $1.24\pm0.08$ & $-0.33\pm0.08$ & VLT & V \\
& 59194.031 & $+2.22$ & $1.35\pm0.09$ & $167.7\pm1.9$ & $1.23\pm0.09$ & $-0.56\pm0.09$ & VLT & V \\
& 59194.038 & $+2.23$ & $1.55\pm0.11$ & $169.8\pm2.0$ & $1.46\pm0.11$ & $-0.54\pm0.11$ & VLT & B \\
& 59194.047 & $+2.24$ & $1.25\pm0.09$ & $159.3\pm2.0$ & $0.94\pm0.09$ & $-0.83\pm0.09$ & VLT & R \\
& 59194.053 & $+2.24$ & $1.24\pm0.09$ & $164.9\pm2.1$ & $1.08\pm0.09$ & $-0.63\pm0.09$ & VLT & V \\
AT2021blz & 59252.038 & $-7.19$ & $0.53\pm0.09$ & $9.9\pm5.0$ & $0.50\pm0.09$ & $0.18\pm0.09$ & VLT & V \\
& 59252.044 & $-7.19$ & $0.54\pm0.10$ & $15.06\pm5.3$ & $0.47\pm0.10$ & $0.27\pm0.10$ & VLT & B \\
& 59263.034 & $+3.80$ & $0.41\pm0.10$ & $16.9\pm7.0$ & $0.34\pm0.10$ & $0.23\pm0.10$ & VLT & V \\
& 59263.040 & $+3.81$ & $<0.35$ & ... & $0.33\pm0.12$ & $-0.06\pm0.09$ & VLT & B \\
& 59263.051 & $+3.82$ & $<1.95$ & ... & $0.33\pm0.12$ & $-0.06\pm0.09$ & VLT & R \\
AT2024pvu & 60545.190 & +1.05 & $<1.42$ & ... & $-0.06\pm0.21$ & $0.26\pm0.21$ & NOT & R \\
 & 60550.076 & +5.93 & $<0.32$ & ... & $0.10\pm0.11$ & $0.18\pm0.11$ & NOT & R\\
 & 60565.142 & +21.00 & $0.38\pm0.12$ & $62.6\pm8.6$ & $-0.24\pm0.12$ & $0.34\pm0.12$ & NOT & R\\
 & 60587.956 & +43.81 & $<0.73$ & ... & $-0.02\pm0.22$ & $-0.02\pm0.21$ & NOT & B\\ 
 & 60587.971 & +43.83 & $<1.06$ & ... & $0.04\pm0.15$ & $0.20\pm0.15$ & NOT & V\\
 & 60587.985 & +43.84 & $<0.36$ & ... & $-0.00\pm0.12$ & $0.33\pm0.12$ & NOT & R\\
 & 60589.024 & +44.88 & $<0.78$ & ... & $0.54\pm0.26$ & $-0.26\pm0.26$ & OSN T90 & R\\
 & 60589.080 & +44.94 & $<0.93$ & ... & $0.50\pm0.31$ & $-0.09\pm0.29$ & OSN T90 & R\\
 & 60591.779 & +47.63 & $0.74\pm0.14$ & $15.2\pm5.4$ & $0.64\pm0.14$ & $0.38\pm0.14$ & NOT & R\\
 & 60592.750 & +48.61 & $<0.44$ & ... & $0.18\pm0.14$ & $-0.02\pm0.19$ & Skinakas 1.3m & r\\
 & 60592.773 & +48.63 & $1.08\pm0.29$ & $37.5\pm7.7$ & $0.28\pm0.29$ & $-1.04\pm0.29$ & Skinakas 1.3m & g\\
 & 60606.940 & +62.80 & $<1.22$ & ... & $-0.37\pm0.42$ & $-0.60\pm0.40$ & Skinakas 1.3m & g\\
 & 60642.732 & +98.59 & $<0.74$ & ... & $0.31\pm0.25$ & $0.03\pm0.23$ & Skinakas 1.3m & r\\
\hline
\end{tabular}
\tablefoot{(1) TDE name. (2) MJD of the observation. (3) Time from peak. (4) Measured polarisation degree, with the associated uncertainty. (5) Measured polarisation angle, with the associated uncertainty. The measured polarisation angle is only reported when the polarisation degree is detected within 3$\sigma$ significance. (6) Measured q Stokes parameter, with the associated uncertainty. (7) Measured u Stokes parameter, with the associated uncertainty. (8) Telescope with which the observation was conducted. The acronyms NOT, VLT and OSN T90 stand for the Nordic Optical Telescope, the Very Large Telescope and the Observatorio de Sierra Nevada 90 cm telescope, respectively. (9) Observing filter.}
\end{table*}

\twocolumn
\section{Fallback time estimates}
\label{polanglefig}

In this section we estimate the fallback time ($t_0$), described as the timescale over which around 50\% of the gas from the tidal disruption is expelled or accreted on the SMBH, based on two different models. In the reprocessing picture we use the black hole mass ($M_{\rm BH}$) and the disrupted stellar mass ($M_*$) inferred with \texttt{MOSFiT} \citep{mockler2019} and adopt Eq.~(5) of \citet{metzger2016}:

\begin{equation}
    t_0 = 41 M_{\bullet,6}^{1/2} M_*^{1/5}\ {\rm d},
\end{equation}where $M_{\bullet,6}$ is the $M_{\rm BH}$ in units of $10^6$ $M_\odot$.  
In the shock-powered emission scenario, $t_0$ is instead estimated from \citep{ryu2020,koljonen2025} as

\begin{equation}
    t_0 = 37 M_{\bullet,6}^{1/2} M_*^{1/3} \Xi^{-3/2} {\rm d},
\end{equation}with the correction factor $\Xi$ accounting for stellar structure and relativistic effects \citep{ryu2020}:

\begin{equation}
    \Xi= \frac{0.62 + \exp[(M_*-0.67)/0.21]}{1+0.55\exp[(M_*-0.67)/0.21]} \left(1.27 - 0.3 M_{\bullet,6}^{0.242}\right).
\end{equation}

We adopted the $M_{\rm BH}$ and $M_*$ values from \citet{floris2025} where available; for sources not included there, we applied the same methodology. In the case of AT2023clx, due to the combination of measured blackbody temperature ($T_{\rm bb}$) and bolometric luminosity ($L_{\rm bol}$) associated with the event, no solution could be found within the allowed parameter range. As a consequence, we assumed the $M_{\rm BH}$ and $M_*$ adopted in \cite{koljonen2025}: $M_{\rm BH}=10^6 M_\odot$, $M_*=0.1 M_\odot$. The mass estimates adopted in this work, together with the estimated $t_0$ values, are reported in Table \ref{tab:masses}.

\begin{table*}[h!]
\renewcommand{\arraystretch}{1.25}
\caption{TDEmass and MOSFit mass results.}
\label{tab:masses}
\centering
\begin{tabular}{l c c c c c c c}
\hline\hline
Name & {\tt TDEMass} $M_{\rm BH}$ &  {\tt TDEMass} $M_*$ & {\tt TDEMass} $t_0$ & {\tt MOSFiT} $M_{\rm BH}$ & {\tt MOSFiT} $M_*$ & {\tt MOSFiT} $t_0$ & Reference\\
 & [$10^6$ M$_\odot$] & [M$_\odot$] & [d] & [$10^6$ M$_\odot$] & [M$_\odot$] & [d] & \\
(1) & (2) & (3) & (4) & (5) & (6) & (7) & (8)\\
\hline
AT2018dyb & $1.60^{+0.18}_{-0.10}$ & $1.20^{+0.05}_{-0.04}$ & 26.0 & $4.07^{+5.05}_{-2.17}$ & $0.98^{+4.56}_{-0.34}$ & 82.0 & This work\\
AT2019dsg & $0.96^{+0.10}_{-0.06}$ & $0.92^{+0.02}_{-0.02}$ & 22.0 & $3.72^{+3.36}_{-1.77}$ & $1.01^{+4.66}_{-0.25}$ & 79.0 & This work\\
AT2019aalc & $8.60^{+0.93}_{-0.92}$ & $1.20^{+0.07}_{-0.06}$ & 80.0 & $26.30^{+26.18}_{-12.81}$ & $1.77^{+8.87}_{-0.92}$ & 73.0 & This work\\
AT2020zso & $6.20^{+0.90}_{-0.82}$ & $0.92^{+0.04}_{-0.04}$ & 50.0 & $1.45^{+1.64}_{-0.84}$ & $0.19^{+1.49}_{-0.13}$ & 35.0 & This work\\
AT2020mot & $2.80^{+0.96}_{-0.78}$ & $0.97^{+0.08}_{-0.07}$ & 42.0 & $18.20^{+15.68}_{-7.88}$ & $0.63^{+2.40}_{-0.56}$ & 159.0 & \cite{floris2025}\\
AT2020afhd & $0.75^{+0.17}_{-0.10}$ & $0.80^{+0.02}_{-0.02}$ & 22.0 & $2.95^{+3.22}_{-1.57}$ & $1.21^{+4.65}_{-0.48}$ & 73.0 & \cite{floris2025}\\ 
AT2021blz & $0.64^{+0.05}_{-0.03}$ & $0.77^{+0.01}_{-0.01}$ & 21.0 & $4.57^{+5.43}_{-2.43}$ & $0.11^{+0.55}_{-0.04}$ & 56.0 & This work\\
AT2022fpx & $10.00^{+0.93}_{-0.89}$  & $0.92^{+0.03}_{-0.03}$ & 108.0 & $213.80^{+132.94}_{-81.97}$ & $14.25^{+51.55}_{-12.42}$ & 1020.0 & \cite{floris2025}\\ 
AT2023clx & ... & ... & 33.0 & $0.47^{+5.29}_{-0.41}$ & $0.05^{+0.30}_{-0.04}$ & 15.0 & \cite{floris2025}\\ 
AT2023lli & $0.58^{+0.40}_{-0.24}$ & $0.51^{+0.07}_{-0.07}$ & 28.0 & $3.47^{+3.29}_{-1.69}$ & $0.40^{+1.64}_{-0.39}$ & 64.0 & \cite{floris2025}\\ 
AT2024gre & $14.00^{+0.51}_{-0.54}$ & $2.10^{+0.22}_{-0.20}$ & 123.0 & $2.51^{+2.86}_{-1.22}$ & $0.57^{+2.31}_{-0.56}$ & 58.0 & \cite{floris2025}\\
AT2024pvu & $8.60^{+0.90}_{-0.87}$ & $1.10^{+0.05}_{-0.04}$ & 84.0 & $13.49^{+10.50}_{-5.90}$ & $0.62^{+2.32}_{-0.61}$ & 137.0 & This work\\
\hline
\end{tabular}
\tablefoot{(1) TNS name of the source. (2) Black hole mass estimated using the {\tt TDEMass} code. (3) Mass of the disrupted star inferred using the {\tt TDEMass} code. (4) Adopted $t_0$ according to the shock model estimated using the mass results from {\tt TDEMass}. (5) Black hole mass estimated using the {\tt MOSFiT} code. (6) Mass of the disrupted star inferred using the {\tt MOSFiT} code. (7) Adopted $t_0$ according to the reprocessing model estimated using the mass results from {\tt MOSFiT}. (8) Reference for these values.}
\end{table*}

\begin{figure}[h!]
\centering
\includegraphics[width=\hsize]{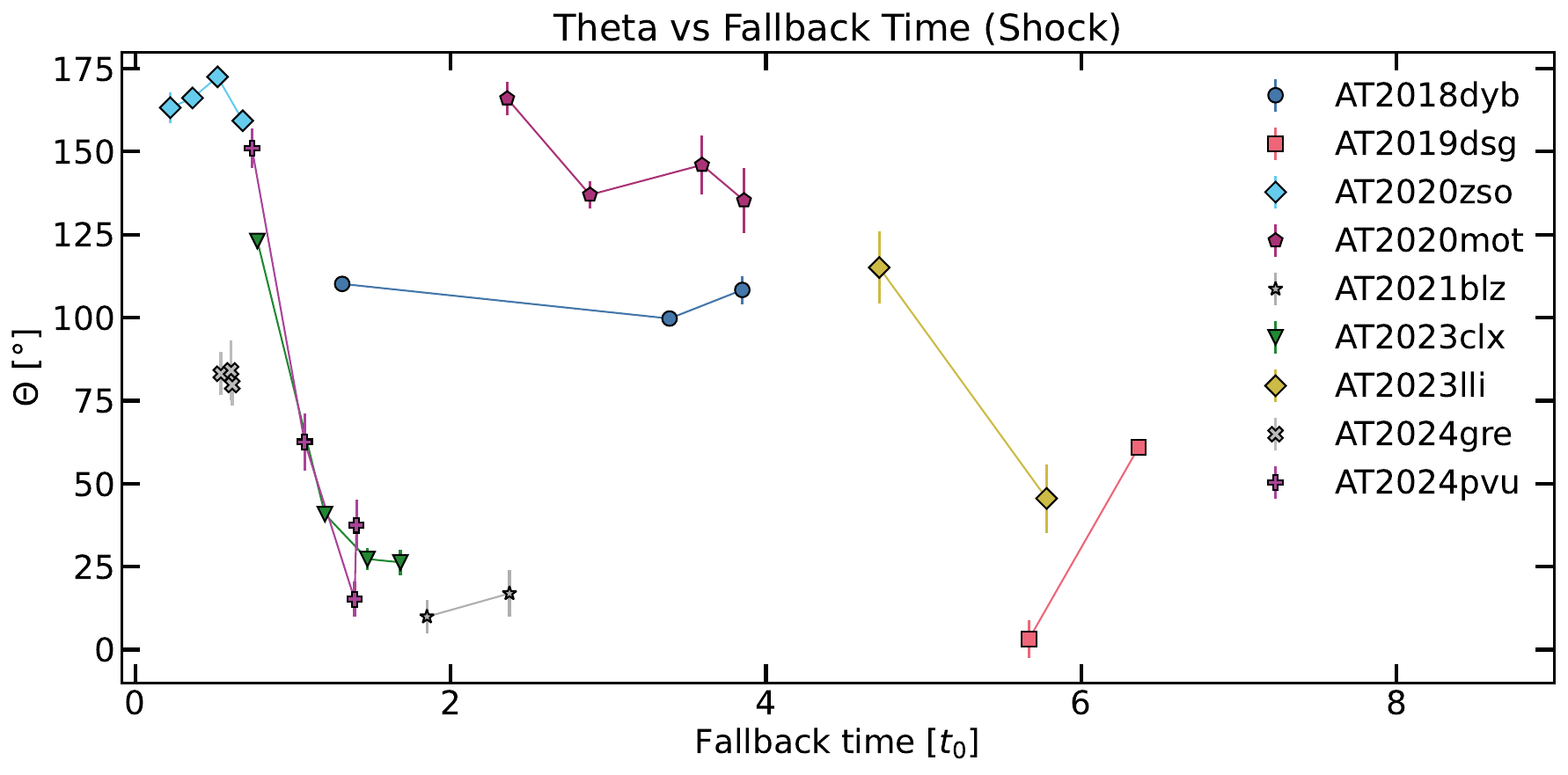}
\caption{$\Theta$ measurements of the sources in the sample, as a function of $t_0$, estimated based on the shock model \citep{ryu2020}. Non-detections are omitted, as $\Theta$ is undefined in such cases.}
\label{fig:bhtimeshocktde}
\end{figure}

\begin{figure}[h!]
\centering
\includegraphics[width=\hsize]{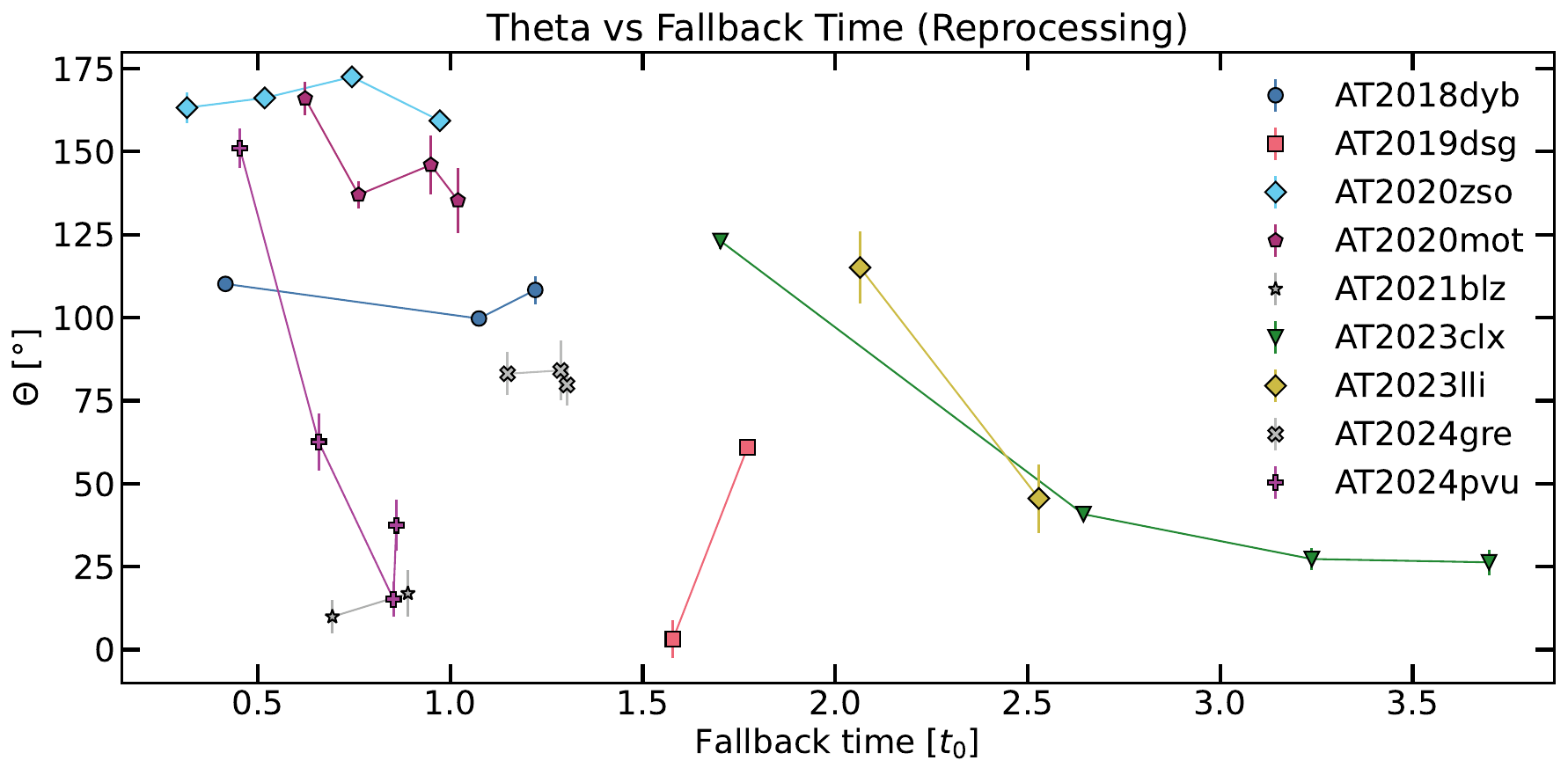}
\caption{$\Theta$ measurements of the sources in the sample, as a function of $t_0$, estimated based on the reprocessing model \citep{mockler2019}. Non-detections are omitted, as $\Theta$ is undefined in such cases.}
\label{fig:bhtimereprtde}
\end{figure}

\begin{figure}[h!]
\centering
\includegraphics[width=\hsize]{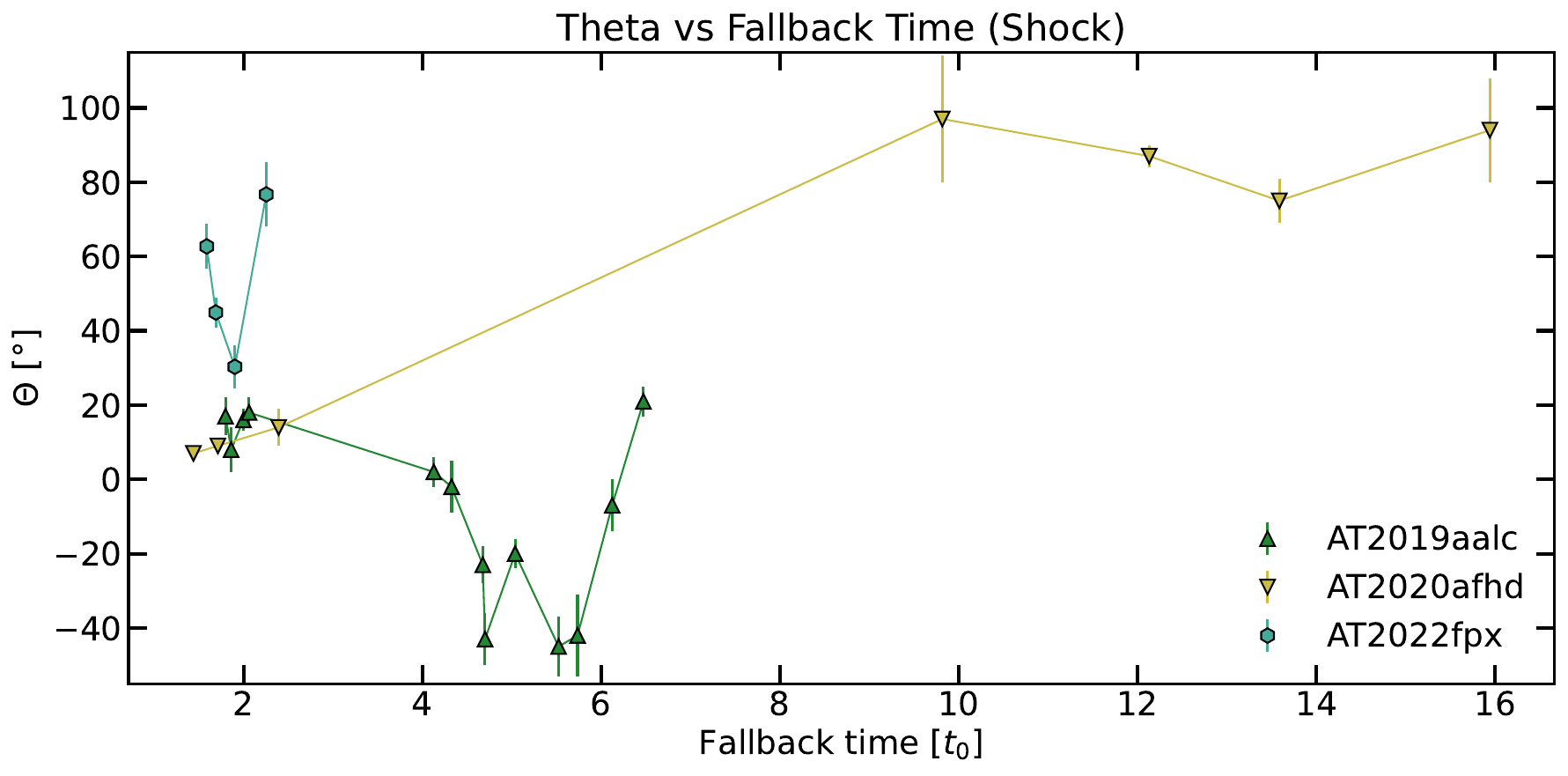}
\caption{$\Theta$ measurements of the sources in the sample, as a function of $t_0$, estimated based on the shock model \citep{ryu2020}. Non-detections are omitted, as $\Theta$ is undefined in such cases.}
\label{fig:bhtimeshockbff}
\end{figure}

\begin{figure}[h!]
\centering
\includegraphics[width=\hsize]{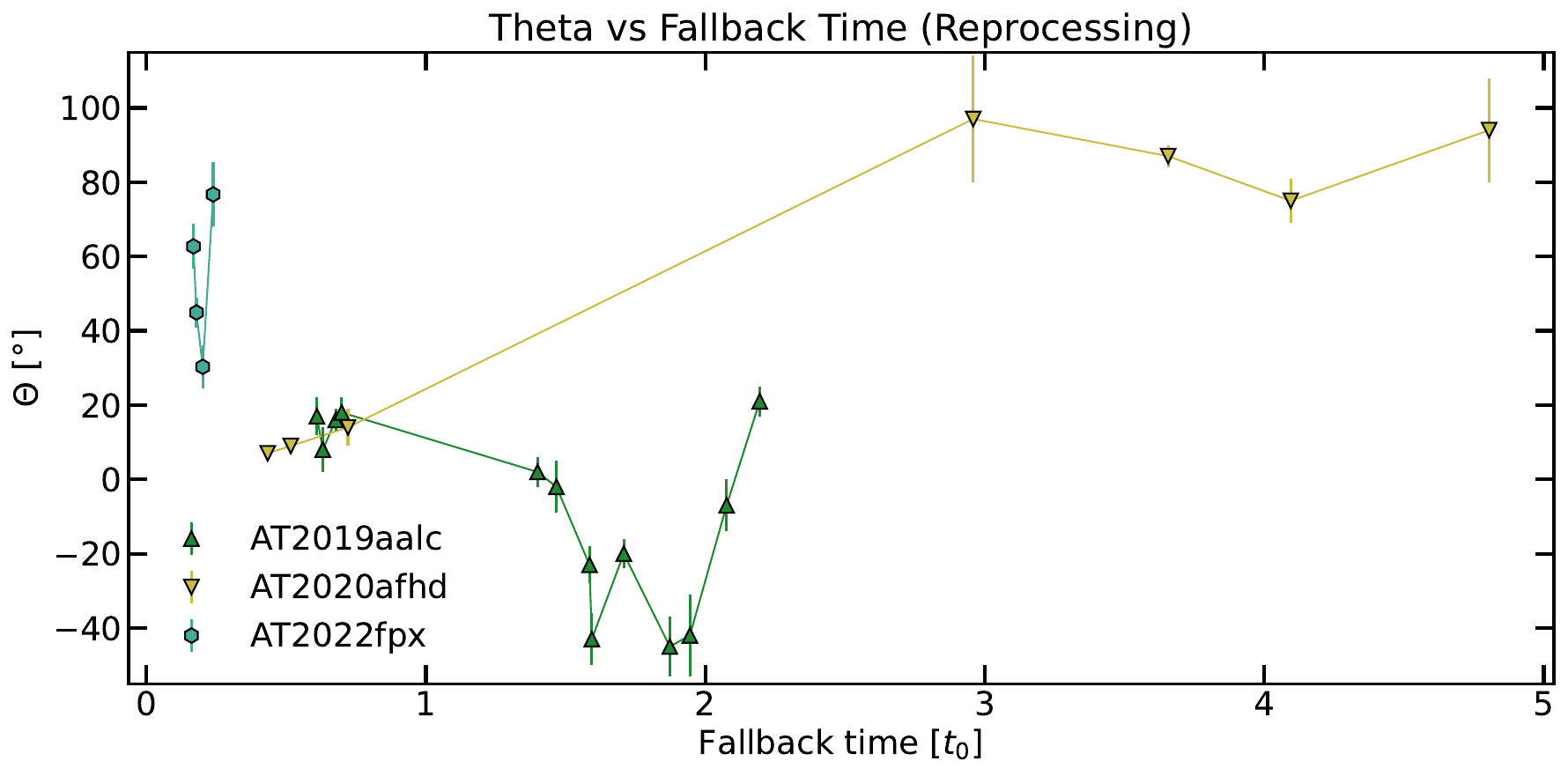}
\caption{$\Theta$ measurements of the sources in the sample, as a function of $t_0$, estimated based on the reprocessing model \citep{mockler2019}. Non-detections are omitted, as $\Theta$ is undefined in such cases.}
\label{fig:bhtimereprbff}
\end{figure}

We display the $\Theta$ variations over time for the simple TDEs according to the $t_0$ from the $M_{\rm BH}$ and $M_*$ results obtained adopting the {\tt TDEMass} shock model (Fig. \ref{fig:bhtimeshocktde}) and the results from the {\tt MOSFiT} reprocessing model (Fig. \ref{fig:bhtimereprtde}). We also display the same for the BFFs alone in Figs. \ref{fig:bhtimeshockbff} and \ref{fig:bhtimereprbff}. We report that no systematic trend was found among the objects in our sample when adopting the $t_0$ time frame, as clearly shown from the figures below. 

\end{appendix}

\end{document}